\documentclass[reprint,aps,prl,superscriptaddress,amsmath,amssymb,floatfix,footinbib,longbibliography]{revtex4-1}
\usepackage{graphicx}
\usepackage{dcolumn}
\usepackage{bm}
\usepackage{amsmath}
\usepackage{times}
\usepackage{color}
\usepackage[breaklinks=true,colorlinks,citecolor=blue,linkcolor=blue,urlcolor=blue]{hyperref}

\makeatletter

\newcommand{\Rmnum}[1]{\expandafter\@slowromancap\romannumeral #1@}
\makeatother

\begin{document}
\title{Angle dependent hysteretic magnetotransport in MnBi$_2$Te$_4$ nanoflakes}
\author{Tithiparna Das}
\affiliation{Department of Physics, Indian Institute of Technology Kanpur, Kanpur 208016, India}
\author{Soumik Mukhopadhyay}
\email{soumikm@iitk.ac.in}
\affiliation{Department of Physics, Indian Institute of Technology Kanpur, Kanpur 208016, India}

\begin{abstract}
Controlling magnetic phases in two-dimensional systems, where charge transport is highly sensitive to real-space spin inhomogeneities, is central to understanding emergent magnetic states in reduced dimensions. In this context, thickness-dependent magnetotransport provides access to irreversible magnetic processes that are not captured by reversible transport or bulk magnetization alone. Here we report an extensive study of hysteretic magnetoresistance in single-crystalline nanoscale thin flakes of the layered antiferromagnet MnBi$_2$Te$_4$. The multi-step hysteresis exhibits a pronounced non-monotonic dependence on thickness and displays nontrivial angular anisotropy. The transport signatures rule out surface-dominated magnetism and simple bulk metamagnetic transitions as the primary origin. We argue that the magnetic irreversibility is possibly governed by domain wall pinning and de-pinning processes within a spatially non-uniform magnetic landscape. These results suggest that reduced dimensionality is a key driver of magnetic irreversibility in MnBi$_2$Te$_4$.
\end{abstract}

\maketitle

{\it Introduction:--} 
Antiferromagnets have emerged as promising platforms for next-generation spintronic and topological devices due to their ultrafast spin dynamics, robustness against external magnetic perturbations, and absence of stray fields \cite{Baltz2018RMPAFMSpintronics, Liu2025MatterAFMMaterials, Zheng2024NatCommunOrbitalTorque}. 
However, unlike ferromagnets, their compensated magnetic structure makes direct detection of internal magnetic rearrangements inherently challenging \cite{Baltz2018RMPAFMSpintronics, Jungwirth2016NatNanoAFM}. As research moves beyond uniform magnetic order and toward field-driven reconfiguration and switching dynamics, identifying reliable experimental signatures of magnetic irreversibility in antiferromagnets has become a central challenge \cite{Stamps2000JPhysDExchangeBias, Radu2008}. In reduced dimensions, this challenge becomes even more pronounced \cite{Marrows2012DomainWalls, Catalan2012RevModPhysNanoelectronics, PhysRevB.111.064304, BoixConstant2024NatMaterSwitching}. When the thickness of a magnetic material approaches intrinsic magnetic length scales, the competition between exchange interaction, magnetic anisotropy, and interlayer coupling is strongly modified by confinement \cite{Catalan2012RevModPhysNanoelectronics, Jaiswal2019JPhysDAnisotropy, nano12234276}. Under these conditions, magnetization reversal need not proceed through simple coherent spin rotation or sharp bulk transitions. Instead, thin magnetic systems often exhibit history-dependent transport, multi-step switching, and nontrivial angular anisotropy \cite{BoixConstant2024NatMaterSwitching, Chhetri2025PRBNegativeMR}. Understanding the microscopic origin of such irreversible behavior is essential for both fundamental physics and device applications \cite{RenukaBalakrishna2024MRSHysteresis}.
Electrical transport particularly provides a powerful probe of magnetic reconfiguration in antiferromagnetic systems \cite{Honda2019PRBUIrSi3, PhysRevB.104.184424}. In materials with strong spin-orbit coupling and topologically nontrivial electronic structure, the electronic response is highly sensitive to the underlying magnetic configuration. As a result, magnetoresistance and Hall measurements can detect subtle changes in magnetic state that remain invisible in conventional magnetometry. However, disentangling whether transport hysteresis originates from surface effects, bulk metamagnetic transitions, coherent spin rotation, or more complex non-uniform magnetic configurations remains an open experimental challenge \cite{Qi2022FrontiersFeNbTe2, Shi2024NanoVDWMagnetism}.

The layered van der Waals antiferromagnetic topological insulator MnBi$_2$Te$_4$ is an ideal platform to explore these issues \cite{Liu2021NatCommunZeroHall, Zhu2021npjCMMnBi2Te4, Rienks2019NatureMnBi2Te4Gap}. It combines A-type antiferromagnetic order with strong uniaxial anisotropy and nontrivial topological electronic states, leading to a pronounced coupling between magnetism and transport \cite{Li2019SciAdvMnBi2Te4, Li2024NatlSciRevMnBi2Te4, He2020npjQuantumMaterMnBi2Te4, Vyazovskaya2025CommMaterMnBi2Te4}. Its thickness can be tuned over a wide range through mechanical exfoliation, enabling access to regimes where finite-size effects and magnetic confinement play a decisive role. While previous studies have revealed field-induced topological phase transitions and anomalous Hall behavior in MnBi$_2$Te$_4$ \cite{Zhu2024NanoLettMnBi2Te4, Guo2024NanoLettMnBiTeCVD, Bac2022npjQMAHEMnBi2Te4, Zverev2023JETPMnBiTeFamily}, the origin of magnetic irreversibility in intermediate thickness regimes remains insufficiently understood.

In this work, we perform angle-dependent magnetotransport study of MnBi$_2$Te$_4$ thin single crystalline flakes of varying thickness $t$ (nm) within the range $ 10 < t < 50 $ to investigate the mechanisms underlying hysteretic transport. By analyzing the evolution of hysteresis area, critical-field structure, and angular anisotropy, we establish experimental constraints on possible microscopic origins of the irreversible response. Rather than assuming a specific magnetic texture, we use the combined thickness and angular dependencies to evaluate competing scenarios and identify the finite-size conditions under which non-uniform magnetic configurations become energetically relevant. Our results demonstrate that reduced dimensionality acts as a key control parameter for magnetic irreversibility in MnBi$_2$Te$_4$ and highlight angular transport as a sensitive probe of magnetic reconfiguration in layered antiferromagnetic topological materials.
\begin{figure*}
\includegraphics[width=0.6\linewidth]{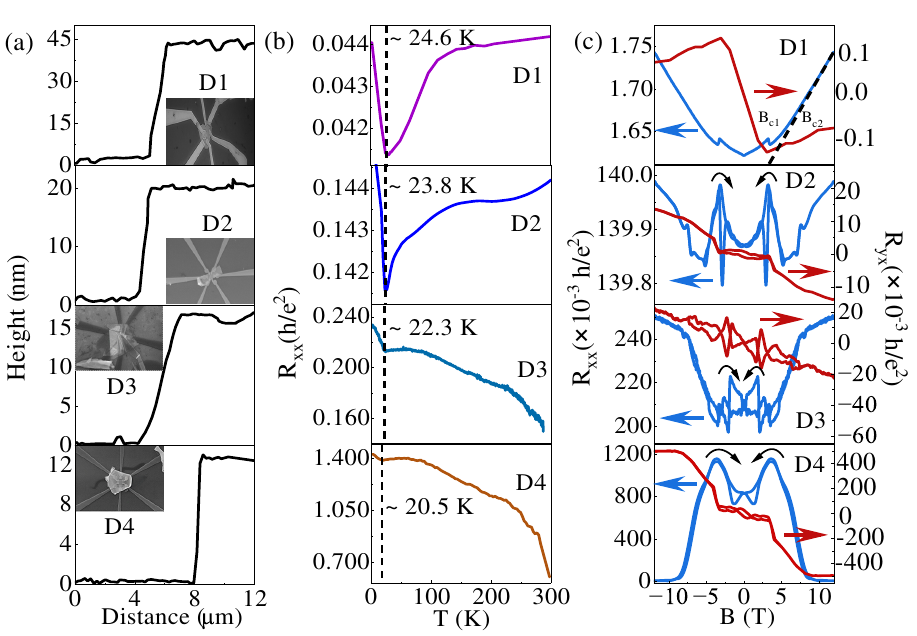}
  \caption{(a) Thickness of prepatterned (D4, D3, D2) and postpatterned (D1) devices measured using the atomic force microscopy (AFM) technique. Inset: SEM images of the devices D1 and D2, D3 and D4 respectively. (b) Temperature dependence of $\mathrm{R_{xx}}$ of MBT samples with different thickness. The dashed lines indicate the Néel temperatures for different devices. (c) Magnetic field dependence of $\mathrm{R_{xx}}$ and $\mathrm{R_{yx}}$ at $\mathrm{T = 2 K}$ along $\mathrm{H || c}$. The black arrows indicate the direction of field sweep.}
  \label{fig1}    
\end{figure*}

{\it Experimental details:--} 
Devices with Hall bar geometry are fabricated using the dry-transfer method. Standard Hall bar electrodes of Ti/Au (15/40 nm) are pre-patterned on a 300 nm oxidized Si/SiO$_2$ substrate via electron beam lithography (EBL), followed by e-beam deposition and a lift-off process. We use scotch tape for the mechanical exfoliation of the bulk crystals of MnBi$_2$Te$_4$ and transfer them onto polydimethylsiloxane (PDMS) stamps. Flakes with thickness $t$ (nm) within the range $ 10 < t < 50 $ are selected and transferred onto the pre-patterned electrodes using an XYZ micromanipulator combined with an optical microscope. The entire transfer process is performed in a glove box under a N$_2$ atmosphere to prevent exposure to oxygen or moisture, with hBN used as a capping layer to protect the devices from H$_2$O and O$_2$. The thicknesses of the fabricated devices are measured using atomic force microscopy (AFM) as shown in Figure~\ref {fig1}(a). For magento-transport measurements, we use a variable temperature insert (VTI) cryostat provided by CRYOGENIC Inc., UK. Longitudinal resistance ($\mathrm{R_{xx}}$) and Hall resistance ($\mathrm{R_{xy}}$) were measured using low-frequency lock-in techniques. Magnetic fields up to 12 T are applied at controlled polar angles ($\mathrm{\theta}$) with respect to the crystallographic easy axis (c-axis), where $\mathrm{\theta = 0^0}$ corresponds to the field along the easy axis. For each thickness, field sweeps are performed in both directions ( $\mathrm{+12 T}$ to $\mathrm{-12 T}$ and vice versa). The raw MR and Hall curves were symmetrized and antisymmetrized, respectively, to eliminate the effects of electrode misalignment.
\begin{figure*}
  \includegraphics[width=0.7\linewidth]{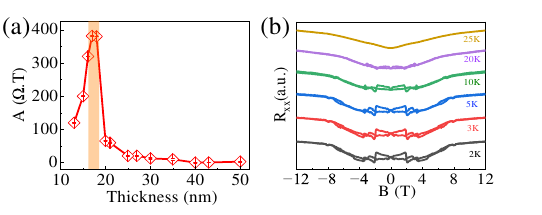}
  \caption{(a) Hysteresis area (A) extracted from $\mathrm{R_{xx}}$ plotted against the thickness of MBT flakes for all devices. The error bars are extracted from the characteristic variation of the resistance difference between the up- and down-sweep curves within the hysteretic field range and propagated to the hysteresis area. The thickness regime with the maximum area has been highlighted. (b) Rxx(B) at different temperatures corresponding to sample thickness indicated in (a).} 
  \label{fig2}
\end{figure*}
\begin{figure*}[htp]
  \includegraphics[width=1\linewidth]{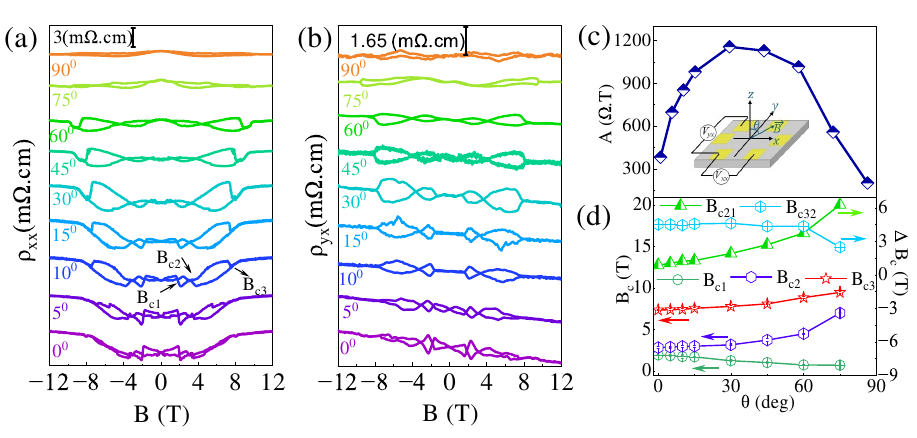}
  \caption{(a) $\mathrm{R_{xx}}$ and (b) $\mathrm{R_{yx}}$ vs magnetic field of D3 plotted at 2 K for different out-of-plane orientations of the external magnetic field. Schematic of the magnetic field rotation is illustrated in the inset of (c). Three critical fields $\mathrm{B_{c1}}$, $\mathrm{B_{c2}}$, and $\mathrm{B_{c3}}$ have been marked by the black arrows as discussed in the text. (c) Hysteresis area plotted as a function of angle of field orientation. (d) Critical fields vs angle of field orientation. The uncertainty in the critical field was estimated from the field interval over which the up- and down-sweep curves merge within the experimental noise level, corresponding to approximately two field steps. }
  \label{fig3}
\end{figure*}

{\it Results and Discussions:--} 
We first discuss the temperature dependence of $\mathrm{R_{xx}}$ of devices with different thicknesses as shown in Figure~\ref {fig1}(b). Device D1 ($\mathrm{\sim 43}$ nm) and D2 ($\mathrm{\sim 20}$ nm) exhibit a distinct change in the slope of $\mathrm{R_{xx}(T)}$ near the Néel temperature ($\mathrm{T_{N}}$). Above $\mathrm{T_{N}}$, $\mathrm{R_{xx}}$ decreases with decreasing temperature, consistent with metallic transport, whereas below $\mathrm{T_{N}}$, the temperature dependence reverses, indicating a transition to insulator-like behaviour. This transition is associated with the development of long-range AFM order, which can alter the scattering landscape and the electronic structure as well, possibly by opening a gap in the surface states. On the other hand, devices D3 ($\mathrm{\sim 17}$ nm) and D4 ($\mathrm{\sim 13}$ nm) exhibit an overall insulating characteristics. These observations suggest that D3 and D4 enter into a regime that is qualitatively different from D1, D2, and the bulk. To get insight into the ground state properties of MBT systematically at different thicknesses, we measure field-dependent $\mathrm{R_{xx}}$ and $\mathrm{R_{xy}}$ at $\mathrm{T = 2}$ K. The four devices show contrasting responses in the presence of magnetic field. Figure~\ref {fig1}(c) summarizes the evolution of magnetotransport hysteresis of all the devices. $\mathrm{R_{xx}}$ of D1 shows surface-dominated linear MR (LMR) followed by two characteristic fields, $\mathrm{B_{c1}}$ and $\mathrm{B_{c2}}$, as discussed in \cite{Lei2020SurfaceLinearMR, Zhu2020PhysRevB101075425}. In device D4, $\mathrm{R_{xy}}$ displays a half-quantized Hall plateau at high magnetic fields, while $\mathrm{R_{xx}}$ simultaneously approaches zero. Such a response is consistent with the realization of a Chern insulator (CI) state hosting two sets of dissipationless chiral edge states as reported earlier \cite{Ge2022PhysRevB105L201404, Liu2020NatMater19_522}. However, the magnetotransport of D2 and D3 is quite different, with a pronounced hysteresis in both $\mathrm{R_{xx}}$ and $\mathrm{R_{xy}}$, and the origin of it remains unexplored to date. In thick films (D1), both $\mathrm{R_{xx}}$ and $\mathrm{R_{xy}}$ are reversible within experimental resolution, showing no detectable hysteresis. Upon reducing the thickness to $\mathrm{\sim 20}$ nm (D2), a clear hysteresis develops in both channels. The hysteresis reaches a pronounced maximum at a thickness of 17–18 nm (D3) and is subsequently suppressed again as the thickness is further reduced to $\mathrm{\sim 13}$ nm (D4). This non-monotonic dependence is highlighted quantitatively in Figure~\ref {fig2}(a), where the total hysteresis area (A) extracted from $\mathrm{R_{xx}}$, is plotted as a function of thickness.

The hysteresis area was calculated by integrating the absolute difference between the symmetrized longitudinal magnetoresistance measured during the up- and down-field sweeps over the magnetic-field range where hysteresis is observed. A similar trend is observed in  $\mathrm{R_{xy}}$ also. The presence of a single maximum around $\mathrm{\sim 18}$ nm rules out surface-dominated magnetism, which would monotonically increase toward thinner films, and is inconsistent with a simple bulk metamagnetic transition. We examine the temperature evolution of $\mathrm{R_{xx}(B)}$ for a representative device (D3), as shown in Figure~\ref {fig2}(b). The hysteresis progressively weakens with increasing temperature and disappears beyond the Néel temperature as expected, confirming the observed irreversibility being associated with the underlying antiferromagnetic order.

To better understand the hysteretic response, we investigate the angle dependence of longitudinal and Hall magnetoresistance, as shown in Figure~\ref{fig3}. The critical fields $\mathrm{B_{c1}}$, $\mathrm{B_{c2}}$, and $\mathrm{B_{c3}}$ are defined as the fields at which successive hysteresis loops in the magnetoresistance close, determined from the convergence of the up and down-sweep traces. Specifically, $\mathrm{B_{c1}}$ marks the closure of the innermost loop, $\mathrm{B_{c2}}$ corresponds to the second loop, and $\mathrm{B_{c3}}$ denotes the field where the outermost loop closes, with the MR becoming fully reversible for $\mathrm{B > B_{c3}}$. Angular magnetotransport measurements reveal a pronounced non-monotonic dependence of the hysteresis in both MR and Hall signals on the field orientation, as shown in Figure~\ref {fig3}(a) and  Figure~\ref {fig3}(b). Starting from the out-of-plane configuration ($\mathrm{\theta = 0^0}$), the hysteresis loops broaden as the magnetic field is tilted away from the c-axis, indicating an extended range of irreversibility. While the hysteresis width increases with angle, the resistance amplitude peaks at a characteristic tilt angle resulting in a maximum hysteresis area around $\mathrm{\theta \sim 30^0}$ (Figure~\ref {fig3}(c)), where the area is defined as the integrated difference between the up- and down-sweep curves. At larger tilt angles the hysteretic response is relatively suppressed.
Further insight is gained from the angular evolution of the critical fields, as shown in Figure~\ref {fig3}(d). The lowest critical field $B_{c1}$ decreases with increasing tilt angle, while the higher critical fields $B_{c2}$ and $B_{c3}$ shift toward larger values. This behavior is reflected in the angle dependence of the field intervals $\Delta B_{c21}=B_{c2}-B_{c1}$ and $\Delta B_{c32}=B_{c3}-B_{c2}$. While $\Delta B_{c21}$ increases monotonically with angle,  $\Delta B_{c32}$ shows the opposite trend, indicating a redistribution of the irreversible response across the magnetic-field range rather than a single magnetic transition. Such contrasting angle dependence of $\Delta B_{c21}$ and $\Delta B_{c32}$ cannot be explained by simple coherent spin rotation or a bulk metamagnetic transition, both of which would be expected to vary monotonically with the projection of the magnetic field along the easy axis. Bulk metamagnetic transition is associated with field-driven spin reorientation. Such transitions typically occur at characteristic fields that depend only weakly on thickness and would be expected to produce a more uniform hysteretic response. Instead, the strong thickness dependence and the presence of multiple field scales with distinct angular dependence point to a more complex evolution of spin configuration involving irreversible processes.

Although the results narrow down on the range of possible microscopic origin of the hysteresis, they do not uniquely identify the underlying mechanism. Several other processes can, in principle, produce hysteretic magnetotransport in magnetic materials, including surface-dominated magnetism, extrinsic measurement effects, and irreversible evolution of non-uniform magnetic states. To identify the dominant mechanism in MnBi$_2$Te$_4$ thin flakes, we now examine these possibilities systematically and assess their consistency with the observed thickness and angular dependencies. 

One possible origin of the hysteresis is surface-dominated magnetism. In this scenario, the hysteretic response would be expected to increase monotonically with decreasing thickness as a result of the growing contribution of surface layers. Instead, we observe a pronounced nonmonotonic thickness dependence, with a maximum around 17–18 nm and a strong suppression in thinner devices. This behavior is inconsistent with a purely surface-controlled mechanism and therefore rules out surface-dominated magnetism as the primary origin. Extrinsic effects such as contact misalignment, thermal drift, or measurement artifacts can also produce apparent hysteresis. However, the hysteretic response is reproducible across multiple devices and persists over repeated magnetic-field sweeps. Moreover, the hysteresis appears only within well-defined ranges of magnetic field and field orientation and disappears smoothly outside these regimes. These systematic trends indicate that the observed hysteresis is intrinsic rather than arising from experimental artifacts. The observations instead suggest that irreversible evolution of non-uniform magnetic configurations under applied magnetic field provides a plausible microscopic origin of the hysteresis. In such a regime, the magnetic configuration does not evolve uniformly but undergoes spatially inhomogeneous reorganization, leading to history-dependent transport behavior. 
\begin{figure}
  \includegraphics[width=1\linewidth]{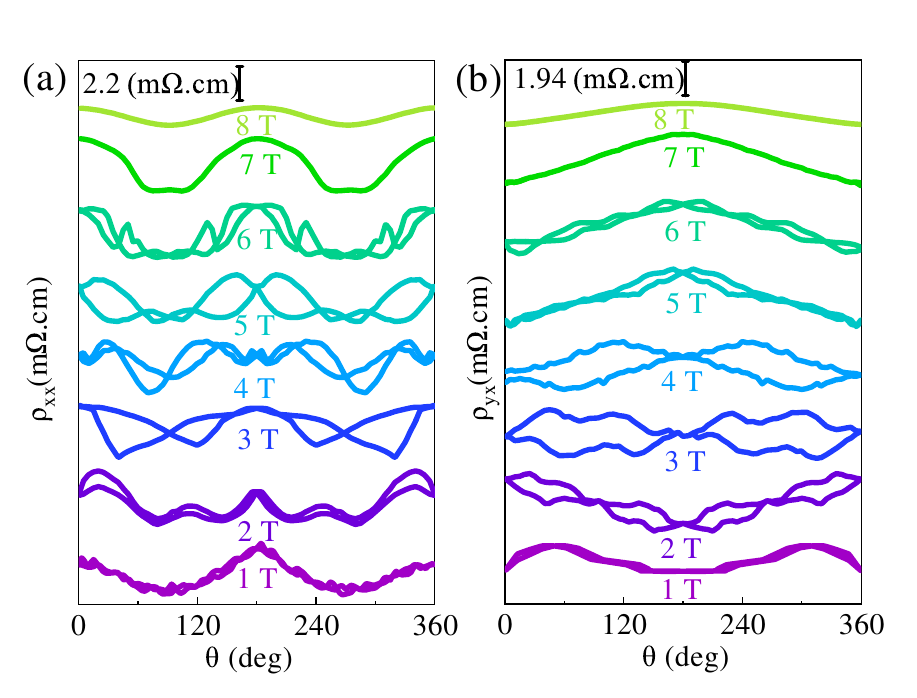}
  \caption{Angle dependent up- and down-sweep  of (a) $\mathrm{\rho_{xx}}$ and (b) $\mathrm{\rho_{yx}}$ at different magnetic fields at $\mathrm{T = 2 K}$.}
  \label{fig4}
\end{figure}
\begin{figure}
  \includegraphics[width=0.8\linewidth]{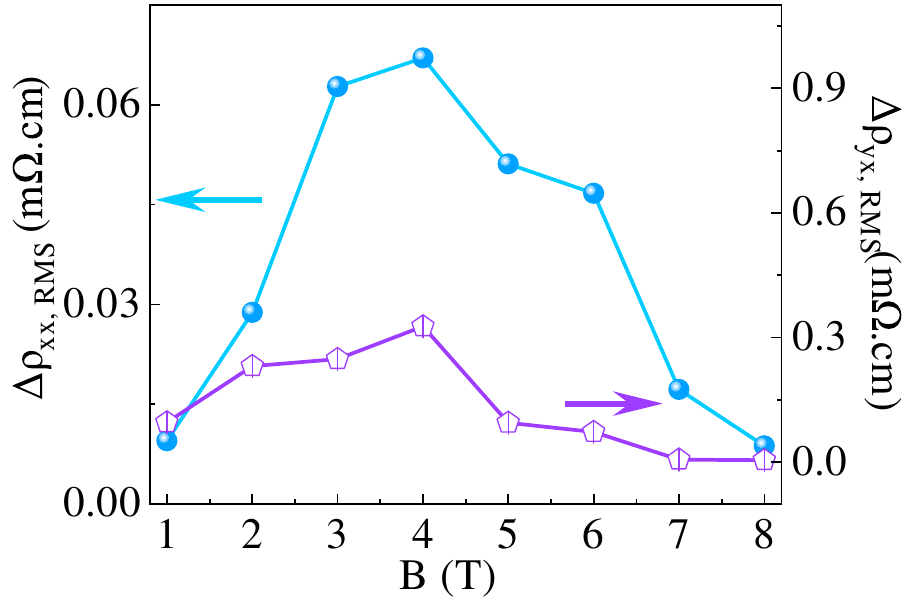}
  \caption{Left panel:  Root-mean-square (RMS) difference between the measured AMR and the expected $\mathrm{cos^2\theta}$ angular dependency with field. Right panel: RMS value of the difference between CW and CCW angular Hall sweeps with field at $\mathrm{T = 2 K}$.}
  \label{fig5}
\end{figure}

Angle-dependent anisotropic magnetoresistance (AMR) measurements in Figure~\ref {fig4}(a) offer additional insight into the nature of the magnetization processes underlying hysteretic transport. In a uniformly magnetized state undergoing coherent rotation, the AMR is expected to follow a simple $\mathrm{cos^2\theta}$ dependence. This behavior is approximately recovered at sufficiently large magnetic fields, where the magnetoresistance becomes fully reversible, and the up- and down-field sweeps coincide, indicating a nearly homogeneous magnetic configuration. In the present case, pronounced deviations from the ideal $\mathrm{cos^2\theta}$ form are observed at intermediate magnetic fields, particularly within the same field range that exhibits strong hysteresis and nontrivial evolution of the critical fields $\mathrm{ B_{c1}}$, $\mathrm{ B_{c2}}$ and $\mathrm{ B_{c3}}$. 

These deviations are quantified by the root-mean-square (RMS) difference between the measured AMR and the expected angular dependence as shown in the left panel of Figure~\ref {fig5}, which shows a clear enhancement in this intermediate-field regime. The confinement of these AMR deviations to a specific field window, rather than their persistence across all fields, indicates that the breakdown of coherent magnetization rotation is not a geometric effect but is instead tied to the underlying magnetic state of the system.  Complementary evidence for angular irreversibility is also provided by the angular Hall measurements. As in Figure~\ref {fig4}(b), at large magnetic fields, the clockwise (CW) and counterclockwise (CCW) angular Hall traces overlap, consistent with reversible magnetization rotation. 

In contrast, a clear separation between CW and CCW sweeps emerges at intermediate magnetic fields, within the same field window where enhanced hysteresis and AMR deviations are observed. This angular Hall hysteresis directly reflects a history-dependent evolution of the out-of-plane magnetization component and cannot be explained within a coherent rotation model. Specifically, the deviation of the AMR from the ideal $\mathrm{cos^2\theta}$ behavior reaches its maximum at the same magnetic field where the RMS value of the difference between CW and CCW angular Hall sweeps is largest as in Figure~\ref {fig5}. 
While the AMR deviation reflects a breakdown of coherent magnetization rotation, the CW–CCW Hall difference directly captures the angular irreversibility of the out-of-plane magnetization component.  The agreement between these two independent angular probes provides strong internal consistency and reinforces the conclusion that the magnetotransport response in this field regime is governed by non-uniform magnetic states.

Among the possible realizations of non-uniform magnetic configurations, domain-wall nucleation and pinning provide a particularly consistent microscopic mechanism for the observed hysteretic magnetotransport in MnBi$_2$Te$_4$. Domain-wall-mediated processes naturally produce history-dependent transport signatures over extended magnetic-field ranges, as well as multi-step critical fields associated with successive nucleation and depinning events. They also provide a natural explanation for the pronounced nonmonotonic thickness dependence observed here, since domain-wall stability and pinning are expected to be optimized when the film thickness becomes comparable to intrinsic magnetic length scales, while thinner films suppress domain formation and thicker films allow more efficient relaxation. At small tilt angles, the dominant out-of-plane field component stabilizes relatively uniform magnetic domains, limiting the number and mobility of domain walls and resulting in modest hysteresis. At intermediate angles, the in-plane field component becomes sufficiently strong to destabilize uniform domains without fully aligning the magnetization, thereby promoting repeated domain-wall nucleation, pinning, and depinning during magnetic-field sweeps. This regime maximizes irreversible domain-wall motion and leads to the largest hysteresis area observed. Upon further increasing the tilt angle, the growing in-plane field component favors smoother magnetic reconfiguration through enhanced domain-wall de-pinning or quasi-coherent rotation, reducing pinning-induced irreversibility and leading to a progressive suppression of the hysteresis at larger angles. The proposed scenario is consistent with recent magnetic imaging studies of MnBi$_2$Te$_4$ \cite{Kim2025AFMTexturesMBT, Sass2020NanoLettMBTDomains, PhysRevMaterials.8.124202}.

{\it Conclusion:--} 
We have systematically investigated hysteretic magnetotransport in MnBi$_2$Te$_4$ thin flakes using magnetic-field sweeps, angular rotation, and thickness-dependent measurements. The observation of multi-step hysteresis with nontrivial angular evolution of the critical fields, together with a nonmonotonic dependence of the hysteresis area on both angle and thickness, reveals a redistribution of irreversibility across the magnetic-field range. Systematic analysis rules out surface-dominated magnetism, simple bulk metamagnetic transitions, coherent rotation and extrinsic effects as the primary origins of the hysteresis. The combined thickness dependence, angular anisotropy, and deviations from coherent rotation instead suggest irreversible evolution of non-uniform magnetic configurations, most consistently explained by domain-wall nucleation and pinning-depinning processes. These results highlight the importance of domain-wall-mediated processes in governing irreversible transport behavior and establish magnetotransport as an alternate, simple and yet sensitive probe of magnetic inhomogeneity in MnBi$_2$Te$_4$ thin flakes. 

{\it Acknowledgements:--} The authors acknowledge IIT Kanpur and the Department of Science and Technology, India, [Order No.
DST/NM/TUE/QM-06/2019 (G)] for financial support. T.D. thanks PMRF for financial support.

\bibliography{ref}

@article{Lei2020SurfaceLinearMR,
  title   = {Surface-induced linear magnetoresistance in the antiferromagnetic topological insulator {MnBi$_2$Te$_4$}},
  author  = {Lei, Xiao and Zhou, L. and Hao, Z. Y. and Ma, X. Z. and Ma, C. and Wang, Y. Q. and Chen, P. B. and Ye, B. C. and Wang, L. and Ye, F. and Wang, J. N. and Mei, J. W. and He, H. T.},
  journal = {Physical Review B},
  volume  = {102},
  number  = {23},
  pages   = {235431},
  year    = {2020},
  doi     = {10.1103/PhysRevB.102.235431},
  url     = {https://doi.org/10.1103/PhysRevB.102.235431}
}

@article{Zhu2020PhysRevB101075425,
  title   = {From negative to positive magnetoresistance in the intrinsic magnetic topological insulator {MnBi$_2$Te$_4$}},
  author  = {Zhu, Peng-Fei and Ye, Xing-Guo and Fang, Jing-Zhi and Xiang, Peng-Zhan and Li, Rong-Rong and Xu, Dai-Yao and Wei, Zhongming and Mei, Jia-Wei and Liu, Song and Yu, Da-Peng and Liao, Zhi-Min},
  journal = {Physical Review B},
  volume  = {101},
  number  = {7},
  pages   = {075425},
  year    = {2020},
  doi     = {10.1103/PhysRevB.101.075425},
  url     = {https://doi.org/10.1103/PhysRevB.101.075425}
}

@article{Zheng2024NatCommunOrbitalTorque,
  author  = {Zheng, Zhenyi and Zeng, Tao and Zhao, Tieyang and Shi, Shu and Ren, Lizhu and Zhang, Tongtong and Jia, Lanxin and Gu, Youdi and Xiao, Rui and Zhou, Hengan and Zhang, Qihan and Lu, Jiaqi and Wang, Guilei and Zhao, Chao and Li, Huihui and Tay, Beng Kang and Chen, Jingsheng},
  title   = {Effective electrical manipulation of a topological antiferromagnet by orbital torques},
  journal = {Nature Communications},
  volume  = {15},
  pages   = {745},
  year    = {2024},
  doi     = {10.1038/s41467-024-45109-1},
  url     = {https://doi.org/10.1038/s41467-024-45109-1}
}

@article{Jungwirth2016NatNanoAFM,
  author  = {Jungwirth, Tom{\' a}{\v{s}} and Marti, Xavier and Wadley, Patrick and Wunderlich, Jairo},
  title   = {Antiferromagnetic spintronics},
  journal = {Nature Nanotechnology},
  volume  = {11},
  pages   = {231--241},
  year    = {2016},
  doi     = {10.1038/nnano.2016.18},
  url     = {https://doi.org/10.1038/nnano.2016.18}
}

@article{Ge2022PhysRevB105L201404,
  title   = {Magnetization-tuned topological quantum phase transition in {MnBi$_2$Te$_4$} devices},
  author  = {Ge, Jun and Liu, Yanzhao and Wang, Pinyuan and Xu, Zhiming and Li, Jiaheng and Li, Hao and Yan, Zihan and Wu, Yang and Xu, Yong and Wang, Jian},
  journal = {Physical Review B},
  volume  = {105},
  number  = {L201404},
  pages   = {L201404},
  year    = {2022},
  doi     = {10.1103/PhysRevB.105.L201404},
  url     = {https://doi.org/10.1103/PhysRevB.105.L201404}
}

@article{Liu2020NatMater19_522,
  title   = {Robust axion insulator and Chern insulator phases in a two-dimensional antiferromagnetic topological insulator},
  author  = {Liu, Chang and Wang, Yongchao and Li, Hao and Wu, Yang and Li, Yaoxin and Li, Jiaheng and He, Ke and Xu, Yong and Zhang, Jinsong and Wang, Yayu},
  journal = {Nature Materials},
  volume  = {19},
  pages   = {522--527},
  year    = {2020},
  doi     = {10.1038/s41563-019-0573-3},
  url     = {https://doi.org/10.1038/s41563-019-0573-3}
}

@article{Li2019SciAdvMnBi2Te4,
  title   = {Intrinsic magnetic topological insulators in van der Waals layered MnBi$_2$Te$_4$-family materials},
  author  = {Li, Jiaheng and Li, Yang and Du, Shiqiao and Wang, Zun and Gu, Bing-Lin and Zhang, Shou-Cheng and He, Ke and Duan, Wenhui and Xu, Yong},
  journal = {Science Advances},
  volume  = {5},
  number  = {6},
  pages   = {eaaw5685},
  year    = {2019},
  doi     = {10.1126/sciadv.aaw5685},
  url     = {https://doi.org/10.1126/sciadv.aaw5685}
}

@article{Li2024NatlSciRevMnBi2Te4,
  title   = {Progress on the antiferromagnetic topological insulator {MnBi$_2$Te$_4$}},
  author  = {Li, Shuai and Liu, Tianyu and Liu, Chang and Wang, Yayu and Lu, Hai-Zhou and Xie, X. C.},
  journal = {National Science Review},
  volume  = {11},
  number  = {2},
  pages   = {nwac296},
  year    = {2024},
  doi     = {10.1093/nsr/nwac296},
  url     = {https://doi.org/10.1093/nsr/nwac296}
}

@article{He2020npjQuantumMaterMnBi2Te4,
  title   = {MnBi$_{2}$Te$_{4}$-family intrinsic magnetic topological materials},
  author  = {He, Ke},
  journal = {npj Quantum Materials},
  volume  = {5},
  pages   = {90},
  year    = {2020},
  doi     = {10.1038/s41535-020-00291-5},
  url     = {https://doi.org/10.1038/s41535-020-00291-5}
}

@article{Vyazovskaya2025CommMaterMnBi2Te4,
  title   = {Intrinsic magnetic topological insulators of the {MnBi$_{2}$Te$_{4}$} family},
  author  = {Vyazovskaya, Alexandra Y. and Bosnar, Mihovil and Chulkov, Evgueni V. and Otrokov, Mikhail M.},
  journal = {Communications Materials},
  volume  = {6},
  pages   = {88},
  year    = {2025},
  doi     = {10.1038/s43246-025-00794-3},
  url     = {https://doi.org/10.1038/s43246-025-00794-3}
}

@article{Liu2021NatCommunZeroHall,
  title   = {Magnetic-field-induced robust zero Hall plateau state in {MnBi$_2$Te$_4$} Chern insulator},
  author  = {Liu, Chang and Wang, Yongchao and Yang, Ming and Mao, Jiahao and Li, Hao and Li, Yaoxin and Li, Jiaheng and Zhu, Haipeng and Wang, Junfeng and Li, Liang and Wu, Yang and Xu, Yong and Zhang, Jinsong and Wang, Yayu},
  journal = {Nature Communications},
  volume  = {12},
  pages   = {4647},
  year    = {2021},
  doi     = {10.1038/s41467-021-25002-x},
  url     = {https://doi.org/10.1038/s41467-021-25002-x}
}

@article{Zhu2021npjCMMnBi2Te4,
  title   = {Tunable dynamical magnetoelectric effect in antiferromagnetic topological insulator {MnBi$_2$Te$_4$} films},
  author  = {Zhu, Tongshuai and Wang, Huaiqiang and Zhang, Haijun and Xing, Dingyu},
  journal = {npj Computational Materials},
  volume  = {7},
  number  = {1},
  pages   = {121},
  year    = {2021},
  doi     = {10.1038/s41524-021-00589-3},
  url     = {https://doi.org/10.1038/s41524-021-00589-3}
}

@article{Rienks2019NatureMnBi2Te4Gap,
  title   = {Large magnetic gap at the Dirac point in {Bi$_2$Te$_3$/MnBi$_2$Te$_4$} heterostructures},
  author  = {Rienks, E. D. L. and Wimmer, S. and S{\'a}nchez-Barriga, J. and Caha, O. and Mandal, P. S. and R{\r u}{\v z}i{\v c}ka, J. and Ney, A. and Steiner, H. and Volobuev, V. V. and Groi{\ss}, H. and Albu, M. and Kothleitner, G. and Michali{\v c}ka, J. and Khan, S. A. and Min{\'a}r, J. and Ebert, H. and Bauer, G. and Freyse, F. and Varykhalov, A. and Rader, O. and Springholz, G.},
  journal = {Nature},
  volume  = {576},
  pages   = {423--428},
  year    = {2019},
  doi     = {10.1038/s41586-019-1826-7},
  url     = {https://doi.org/10.1038/s41586-019-1826-7}
}

@article{Baltz2018RMPAFMSpintronics,
  title   = {Antiferromagnetic spintronics},
  author  = {Baltz, Vincent and Manchon, Aur{\'e}lien and Tsoi, Mikhail and Moriyama, Takahiro and Ono, Teruo and Tserkovnyak, Yaroslav},
  journal = {Reviews of Modern Physics},
  volume  = {90},
  number  = {1},
  pages   = {015005},
  year    = {2018},
  doi     = {10.1103/RevModPhys.90.015005},
  url     = {https://link.aps.org/doi/10.1103/RevModPhys.90.015005}
}

@article{Liu2025MatterAFMMaterials,
  title   = {Antiferromagnetic materials: From fundamentals to applications},
  author  = {Liu, Jiahao and Lu, Jiaqi and Peng, Shouzhong and Liu, Zhaochun and Zhang, Yongzhuo and Qiao, Jun and Wang, Shuo and Li, Weixiang and Chen, Jingsheng and Wang, Zhiming and Li, Run-Wei and Zhang, Yue and Zhao, Weisheng},
  journal = {Matter},
  volume  = {8},
  number  = {11},
  pages   = {102472},
  year    = {2025},
  doi     = {10.1016/j.matt.2025.102472},
  url     = {https://doi.org/10.1016/j.matt.2025.102472}
}

@article{Stamps2000JPhysDExchangeBias,
  title   = {Mechanisms for exchange bias},
  author  = {Stamps, R. L.},
  journal = {Journal of Physics D: Applied Physics},
  volume  = {33},
  number  = {23},
  pages   = {R247--R268},
  year    = {2000},
  doi     = {10.1088/0022-3727/33/23/201},
  url     = {https://doi.org/10.1088/0022-3727/33/23/201}
}

@Inbook{Radu2008,
title  =  {Exchange Bias Effect of Ferro-/Antiferromagnetic Heterostructures},
bookTitle  =  {Magnetic Heterostructures: Advances and Perspectives in Spinstructures and Spintransport},
author  =  {Radu, Florin and Zabel, Hartmut},
year = {2008},
publisher = {Springer Berlin Heidelberg},
pages = {97--184},
isbn = {978-3-540-73462-8},
doi = {10.1007/978-3-540-73462-8_3},
url = {https://doi.org/10.1007/978-3-540-73462-8_3}
}

@article{Marrows2012DomainWalls,
  author  = {Marrows, C. H. and Meier, G.},
  title   = {Domain wall dynamics in nanostructures},
  journal = {Journal of Physics: Condensed Matter},
  volume  = {24},
  number  = {2},
  pages   = {020301},
  year    = {2012},
  doi     = {10.1088/0953-8984/24/2/020301},
  url     = {https://doi.org/10.1088/0953-8984/24/2/020301}
}

@article{Catalan2012RevModPhysNanoelectronics,
  title   = {Domain wall nanoelectronics},
  author  = {Catalan, G. and Seidel, J. and Ramesh, R. and Scott, J. F.},
  journal = {Reviews of Modern Physics},
  volume  = {84},
  number  = {1},
  pages   = {119--156},
  year    = {2012},
  doi     = {10.1103/RevModPhys.84.119},
  url     = {https://doi.org/10.1103/RevModPhys.84.119}
}

@article{Jaiswal2019JPhysDAnisotropy,
  title   = {Tuning of interfacial perpendicular magnetic anisotropy and domain structures in magnetic thin film multilayers},
  author  = {Jaiswal, Samridh and Lee, K. and Langer, J{\"u}rgen and Ocker, Berthold and Kl{\"a}ui, Mathias and Jakob, Gerhard},
  journal = {Journal of Physics D: Applied Physics},
  volume  = {52},
  number  = {29},
  pages   = {295001},
  year    = {2019},
  doi     = {10.1088/1361-6463/ab1c42},
  url     = {https://doi.org/10.1088/1361-6463/ab1c42}
}

@article{PhysRevB.111.064304,
  title = {Hysteresis loop shape of field-driven antiferromagnetic-ferromagnetic bilayers in experimental conditions},
  author = {Mijatovi\ifmmode \acute{c}\else \'{c}\fi{}, Svetislav and Graovac, Stefan and Spasojevi\ifmmode \acute{c}\else \'{c}\fi{}, Djordje and Tadi\ifmmode \acute{c}\else \'{c}\fi{}, Bosiljka},
  journal = {Phys. Rev. B},
  volume = {111},
  issue = {6},
  pages = {064304},
  numpages = {9},
  year = {2025},
  month = {Feb},
  publisher = {American Physical Society},
  doi = {10.1103/PhysRevB.111.064304},
  url = {https://link.aps.org/doi/10.1103/PhysRevB.111.064304}
}

@article{BoixConstant2024NatMaterSwitching,
  author  = {Boix-Constant, Carla and Jenkins, Sarah and Rama-Eiroa, Ricardo and Santos, Elton J. G. and Ma{\~n}as-Valero, Samuel and Coronado, Eugenio},
  title   = {Multistep magnetization switching in orthogonally twisted ferromagnetic monolayers},
  journal = {Nature Materials},
  volume  = {23},
  number  = {2},
  pages   = {212--218},
  year    = {2024},
  doi     = {10.1038/s41563-023-01735-6},
  url     = {https://doi.org/10.1038/s41563-023-01735-6},
  issn    = {1476-4660}
}

@article{nano12234276,
  author  = {Zsurzsa, S{\'a}ndor and El-Tahawy, Moustafa and P{\'e}ter, L{\'a}szl{\'o} and Kiss, L{\'a}szl{\'o} Ferenc and Gubicza, Jen{\H{o}} and Moln{\'a}r, Gy{\"o}rgy and Bakonyi, Imre},
  title   = {Spacer Layer Thickness Dependence of the Giant Magnetoresistance in Electrodeposited Ni-Co/Cu Multilayers},
  journal = {Nanomaterials},
  volume  = {12},
  number  = {23},
  pages   = {4276},
  year    = {2022},
  doi     = {10.3390/nano12234276},
  url     = {https://doi.org/10.3390/nano12234276},
  issn    = {2079-4991},
  pmid    = {36500898}
}

@article{Chhetri2025PRBNegativeMR,
  title   = {Large negative magnetoresistance in an antiferromagnet},
  author  = {Karki Chhetri, S. and others},
  journal = {Physical Review B},
  volume  = {111},
  pages   = {014431},
  year    = {2025},
  doi     = {10.1103/PhysRevB.111.014431},
  url     = {https://doi.org/10.1103/PhysRevB.111.014431}
}

@article{RenukaBalakrishna2024MRSHysteresis,
  author  = {Renuka Balakrishna, Ananya},
  title   = {Rethinking hysteresis in magnetic materials},
  journal = {MRS Communications},
  volume  = {14},
  number  = {5},
  pages   = {835--845},
  year    = {2024},
  doi     = {10.1557/s43579-024-00624-6},
  url     = {https://doi.org/10.1557/s43579-024-00624-6},
  issn    = {2159-6867}
}

@article{Honda2019PRBUIrSi3,
  title   = {Magnetotransport as a probe of phase transformations in metallic antiferromagnets: The case of {UIrSi$_3$}},
  author  = {Honda, Fuminori and Valenta, Jaroslav and Prokle{\v s}ka, Ji{\v r}{\'\i} and Posp{\'\i}{\v s}il, Ji{\v r}{\'\i} and Proschek, Petr and Prchal, Ji{\v r}{\'\i} and Sechovsk{\'y}, Vladim{\'\i}r},
  journal = {Physical Review B},
  volume  = {100},
  number  = {1},
  pages   = {014401},
  year    = {2019},
  doi     = {10.1103/PhysRevB.100.014401},
  url     = {https://doi.org/10.1103/PhysRevB.100.014401}
}

@article{PhysRevB.104.184424,
  title = {Hysteretic effects and magnetotransport of electrically switched {CuMnAs}},
  author = {Zub\'a\ifmmode \check{c}\else \v{c}\fi{}, Jan and Ka\ifmmode \check{s}\else \v{s}\fi{}par, Zden\ifmmode \check{e}\else \v{e}\fi{}k and Krizek, Filip and F\"orster, Tobias and Campion, Richard P. and Nov\'ak, V\'{\i}t and Jungwirth, Tom\'a\ifmmode \check{s}\else \v{s}\fi{} and Olejn\'{\i}k, Kamil},
  journal = {Phys. Rev. B},
  volume = {104},
  issue = {18},
  pages = {184424},
  numpages = {7},
  year = {2021},
  month = {Nov},
  publisher = {American Physical Society},
  doi = {10.1103/PhysRevB.104.184424},
  url = {https://link.aps.org/doi/10.1103/PhysRevB.104.184424}
}

@article{Qi2022FrontiersFeNbTe2,
  title   = {Abnormal Magnetoresistance Transport Properties of van der Waals Antiferromagnetic {FeNbTe$_2$}},
  author  = {Qi, Bingtian and Guo, Junjie and Miao, Yaqi and Zhong, Mingzhe and Li, Bin and Luo, Zeyu and Wang, Xiangguo and Nie, Yuzhe and Xia, Qilin and Guo, Guanghui},
  journal = {Frontiers in Physics},
  volume  = {10},
  pages   = {851838},
  year    = {2022},
  doi     = {10.3389/fphy.2022.851838},
  url     = {https://doi.org/10.3389/fphy.2022.851838}
}

@article{Shi2024NanoVDWMagnetism,
  title   = {Recent Progress in Two-Dimensional Magnetic Materials},
  author  = {Shi, Guang and others},
  journal = {Nanomaterials},
  volume  = {14},
  number  = {21},
  pages   = {1759},
  year    = {2024},
  doi     = {10.3390/nano14211759},
  url     = {https://doi.org/10.3390/nano14211759}
}

@article{Zhu2024NanoLettMnBi2Te4,
  title   = {Unveiling the Anomalous Hall Response of the Magnetic Structure Changes in the Epitaxial {MnBi$_2$Te$_4$} Films},
  author  = {Zhu, Kai and Cheng, Y. and Liao, M. and Chong, S. K. and Zhang, D. and He, K. and Wang, K. L. and Chang, K. and Deng, P.},
  journal = {Nano Letters},
  volume  = {24},
  number  = {7},
  pages   = {2181--2187},
  year    = {2024},
  doi     = {10.1021/acs.nanolett.3c04095},
  url     = {https://doi.org/10.1021/acs.nanolett.3c04095},
  pmid    = {38340079},
  pmcid   = {PMC10885191}
}

@article{Guo2024NanoLettMnBiTeCVD,
  author  = {Guo, Hui and Bai, Chenyu and Zhu, Ke and Lv, Senhao and Zhai, Zhaoyi and Qu, Jingyuan and Xian, Guoyu and Han, Yechao and Hu, Guojing and Qi, Qi and Liu, Guangtong and Jiao, Fang and Bao, Lihong and Bao, Xiaotian and Liu, Xinfeng and Chen, Hui and Lin, Xiao and Zhou, Wu and Zhou, Jiadong and Yang, Haitao and Gao, Hong-Jun},
  title   = {Controllable Synthesis of High-Quality Magnetic Topological Insulator {MnBi$_2$Te$_4$} and {MnBi$_4$Te$_7$} Multilayers by Chemical Vapor Deposition},
  journal = {Nano Letters},
  volume  = {24},
  number  = {49},
  pages   = {15788--15795},
  year    = {2024},
  doi     = {10.1021/acs.nanolett.4c04700},
  url     = {https://doi.org/10.1021/acs.nanolett.4c04700}
}

@article{Bac2022npjQMAHEMnBi2Te4,
  author  = {Bac, S.-K. and Koller, K. and Lux, F. and Wang, J. and Riney, L. and Borisiak, K. and Powers, W. and Zhukovskyi, M. and Orlova, T. and Dobrowolska, M. and Furdyna, J. K. and Dilley, N. R. and Rokhinson, L. P. and Mokrousov, Y. and McQueeney, R. J. and Heinonen, O. and Liu, X. and Assaf, B. A.},
  title   = {Topological response of the anomalous Hall effect in {MnBi$_2$Te$_4$} due to magnetic canting},
  journal = {npj Quantum Materials},
  volume  = {7},
  pages   = {46},
  year    = {2022},
  doi     = {10.1038/s41535-022-00455-5},
  url     = {https://doi.org/10.1038/s41535-022-00455-5}
}

@article{Zverev2023JETPMnBiTeFamily,
  author  = {Zverev, V. N. and Abdullayev, N. A. and Aliyev, Z. S. and Amiraslanov, I. R. and Otrokov, M. M. and Mamedov, N. T. and Chulkov, E. V.},
  title   = {Transport Properties of the Magnetic Topological Insulators Family {(MnBi$_2$Te$_4$)(Bi$_2$Te$_3$)$_m$} ($m = 0, 1, \ldots, 6$)},
  journal = {JETP Letters},
  volume  = {118},
  number  = {12},
  pages   = {905--910},
  year    = {2023},
  month   = {Dec},
  doi     = {10.1134/S0021364023603305},
  url     = {https://doi.org/10.1134/S0021364023603305},
  issn    = {1090-6487}
}

@article{Kim2025AFMTexturesMBT,
  author  = {Kim, Min Gyu and Boney, Starr and Burgard, Luke and Rutowski, Lillian and Mazzoli, Claudio},
  title   = {Dynamics and Formation of Antiferromagnetic Textures in MnBi$_2$Te$_4$ Single Crystal},
  journal = {Materials},
  volume  = {18},
  number  = {23},
  pages   = {5337},
  year    = {2025},
  doi     = {10.3390/ma18235337},
  url     = {https://doi.org/10.3390/ma18235337}
}

@article{Sass2020NanoLettMBTDomains,
  author  = {Sass, Paul M. and Ge, Wenbo and Yan, Jiaqiang and Obeysekera, D. and Yang, J. J. and Wu, Weida},
  title   = {Magnetic Imaging of Domain Walls in the Antiferromagnetic Topological Insulator MnBi$_2$Te$_4$},
  journal = {Nano Letters},
  volume  = {20},
  number  = {4},
  pages   = {2609--2614},
  year    = {2020},
  doi     = {10.1021/acs.nanolett.0c00114},
  url     = {https://doi.org/10.1021/acs.nanolett.0c00114}
}

@article{PhysRevMaterials.8.124202,
  title = {Correlation between magnetic domain structures and quantum anomalous Hall effect in epitaxial $\mathrm{Mn}{\mathrm{Bi}}_{2}{\mathrm{Te}}_{4}$ thin films},
  author = {Shi, Yang and Bai, Yunhe and Li, Yuanzhao and Feng, Yang and Li, Qiang and Zhang, Huanyu and Chen, Yang and Tong, Yitian and Luan, Jianli and Liu, Ruixuan and Ji, Pengfei and Gao, Zongwei and Guo, Hangwen and Zhang, Jinsong and Wang, Yayu and Feng, Xiao and He, Ke and Zhou, Xiaodong and Shen, Jian},
  journal = {Phys. Rev. Mater.},
  volume = {8},
  issue = {12},
  pages = {124202},
  numpages = {7},
  year = {2024},
  month = {Dec},
  publisher = {American Physical Society},
  doi = {10.1103/PhysRevMaterials.8.124202},
  url = {https://link.aps.org/doi/10.1103/PhysRevMaterials.8.124202}
}
\end{document}